\documentclass[aps]{revtex4}
\usepackage{graphicx}
\begin{document}
\title{A sign change of the moment  at $n=0$ of the polarized structure
function $g_1^{p}$}
\author{Susumu Koretune}
\affiliation{Department of Physics, Shimane University,
Matsue,Shimane,690-8504,Japan}
\begin{abstract}The sum rule for the structure function $g_1^{p}$ which
is related to the cross section of the photo-production is used to
show that a sign change of the integral corresponding to the
appropriately defined moment at $n=0$ where the integral is cut
at the point $x=x_c(Q^2)$ occurs at very small $Q^2$
near $Q^2\sim 0.09$(GeV/c)$^2$ and $x_c\sim 0.024$.
This fact shows that the origin of the sign difference between the
Ellis-Jaffe sum rule and the Drell-Hearn-Gerasimov sum rule
lies in the rapid change of the elastic contribution
at low $Q^2$ which is compensated by the inelastic contribution
to satisfy the sum rule at $n=0$. Hence it occurs at very small
$Q^2$.       
\end{abstract}
\maketitle
The fact that the sign of the Drell-Hearn-Gerasimov 
sum rule\cite{Drell,Ger} and that of the Ellis-Jaffe sum
rule\cite{Ellis} was different had motivated
the study of these sum rules and the spin structure functions $g_1$ and $g_2$
at low $Q^2$ from both the experiment and the theory\cite{Xi,Bur,Dres}. 
The Drell-Hearn-Gerasimov (DHG) sum rule stands on the sound
theoretical footings. It is based on the general principles such as 
causality and unitarity.  
However, since this sum rule holds only at $Q^2=0$, 
a theoretical framework which can treat the region $Q^2\neq 0$ 
in a non-perturbative way with a similar generality as 
the DHG sum rule is necessary to study the origin of the sign change.
Many years ago, the sum rules for the structure functions $g_1^{p}$ 
and $g_1^{n}$ which were related to the cross sections of the 
photo-production were derived\cite{kore}. These sum rules were
based on the general principles as in the case of the
DHG sum rule but corresponded to the moment at $n=0$ of 
the polarized structure 
functions $g_1$ and $g_2$. Hence, compared with the DHG
sum rule, they depended more on the high energy behavior of the structure
functions and the cross sections of the photo-production. In this paper,
using the phenomenological study of the high energy behavior of
these quantity\cite{Sim}, we transform the sum rules which are 
heavily related to the behavior 
in the very high energy region into the ones at low energy 
which can be accessible by the experiment, and show that 
at very small $Q^2$ there is a sign
change of the appropriately defined moment at $n=0$ of the structure
function $g_1^p$.\\

Let us first briefly explain how the sum rules can be obtained.
The Deser-Gilbert-Sudarshan(DGS) representation which incorporates
both causality and the spectrum conditions for the hadron has been
of great value in the investigation of the one-particle connected
matrix element of the current commutator\cite{DGS,Con}.
If the lowest mass $M_s$ in the $s$ channel and that of the $M_u$
in the $u$ channel satisfys the condition $m\leq (M_s + M_u)/2$ where $m$ is
the mass of the hadron of the one-particle state, 
this representation can be generalized to the product
of the currents hence to the anticommutation relation of them. 
The stable hadron such as the nucleon satisfys the spectral condition.
Then, we can consider the restriction of the current commutator
and the current anticommutator to the null-plane with the same
weight functions in the DGS representation, and using
information of the current commutation relation based on the
canonical quantization on the null-plane,
we see that which terms in the DGS representation remain 
at the null-plane\cite{kore80,kore84,kore93}. 
In this way, we find that the anticommutaion relation on the 
null-plane becomes
\begin{eqnarray}
\lefteqn{<p|\{J_a^+(x),J_b^i(0)\}|p>_c\delta(x^+) } \nonumber \\
&=&<p|\{s^{+\beta
 i\alpha}\partial_{\alpha}[\Delta^{(1)}(x)G_{c\beta}(x|0)]
-2g^{+\alpha}g^{i\beta}[\partial_{\alpha}[\Delta^{(1)}(x)]G_{c\beta}(x|0)\nonumber \\
&-&\epsilon^{+i\alpha\beta}\partial_{\alpha}[\Delta^{(1)}(x)G_{c\beta}^5(x|0)]\}|p>_c,
\end{eqnarray}
where $\Delta^{(1)}(x)$ and $\partial^{+}\Delta^{(1)}(x)$ read
$(-1/2\pi)\ln |x^{-}|\delta (\vec{x}^{\bot})\delta (x^{+})$ and\\
$(-1/2\pi)P(1/x^{-})\delta (\vec{x}^{\bot})\delta (x^{+})$
respectively, and
$G_c^{\beta}(x|0)$ and $G_c^{5\beta}(x|0)$ are decomposed as
the symmetric and the antisymmetric combinations as
\begin{equation} 
G_c^{\beta}(x|0)=d_{abc}A_c^{\beta}(x|0)+f_{abc}S_c^{\beta}(x|0),
\end{equation}
\begin{equation}
G_c^{5\beta}(x|0)=d_{abc}S_c^{5\beta}(x|0)-f_{abc}A_c^{5\beta}(x|0).
\end{equation}
Then the connected matrix elements are defined as
\begin{equation}
<p|S_a^{\mu}(x|0)|p>_c=p^{\mu}S_a(p\cdot x , x^2) +
x^{\mu}\bar{S}_a(p\cdot x , x^2) ,
\end{equation}
\begin{equation}
<p|S_a^{5\mu}(x|0)|p>_c=s^{\mu}S_c^{5}(p\cdot x,x^2)+p^{\mu}\bar{S}_a(p\cdot x,x^2) +
x^{\mu}(x\cdot s)\tilde{S}_a(p\cdot x,x^2) ,
\end{equation}
where $2s^{\mu}=\bar{u}(p)\gamma^{\mu}\gamma^5u(p)$
with $s\cdot p=0, s^2=-1$ and we set $m_{N}=1$ for simplicity and
similar definitions for the antisymmetric bilocal quantities.
Intuitively, the bilocal quantity in Eqs.(2) and (3) can
be interpreted as the bilocal currents constructed by the quark
bilinear. However, it should be noted that these quantities are defined
only as the connected one-particle matrix elements as
given on the right side of Eqs.(4) and (5), hence we need 
no explicit form of the bilocal quantities for 
the derivation of the sum rule.

The antisymmetric part of the hadronic tensor for the
electromagnetic current is defined as
\begin{eqnarray}
\widetilde{W}_{\mu\nu}^{ab}&=&\frac{1}{4\pi}\int d^4x\exp (iqx)
<p,s|\{J_{\mu}^a(x),J_{\nu}^b(0)\}|p,s>_c \nonumber \\
&=&i\epsilon_{\mu\nu\lambda\sigma}q^{\lambda}s^{\sigma}\tilde{G}_1^{ab}
+i\epsilon_{\mu\nu\lambda\sigma}q^{\lambda}(\nu s^{\sigma}-
q\cdot sp^{\sigma})\tilde{G}_2^{ab}.
\end{eqnarray}
The structure function $\tilde{G}_i$ for $i=1,2$ has opposite
crossing property under $q\to -q,a\leftrightarrow b$ and
$\mu\leftrightarrow \nu$ compared with the usual one
defined by the current commutation relation. Now following
the standard method to get the fixed-mass sum rule in the null-plane
formalism\cite{DJT}, we obtain the two sum rules
\begin{eqnarray}
\int_0^1\frac{dx}{x}g_1^{ab}(x,Q^2)&=&
\frac{-1}{8\pi}d_{abc}\int_{-\infty}^{\infty}d\alpha
\ln |\alpha |[S_c^5(\alpha ,0)+\alpha \bar{S}_c^5(\alpha ,0)], \\
\int_0^1\frac{dx}{x}g_2^{ab}(x,Q^2)&=&
\frac{1}{8\pi}d_{abc}\int_{-\infty}^{\infty}d\alpha
\alpha \ln |\alpha |\bar{S}_c^5(\alpha ,0),
\end{eqnarray}
where we set $\nu =p\cdot q, \alpha =p\cdot x, -q^2=\vec{q}_{\bot}^2=Q^2$, and use the fact
that ,in the $s$ channel ,$\tilde{G}_i$ is the same as the structure function $G_i$ 
defined by the current commutation relation and
that $g_1^{ab}=\nu G_1^{ab}$ and $g_2^{ab}=\nu^2 G_2^{ab}$.
Since the right side of Eqs.(7) and (8) is $Q^2$ independent, we obtain
\begin{equation}
\int_0^1\frac{dx}{x}g_1(x,Q^2)=\int_0^1\frac{dx}{x}g_1(x,Q_0^2)
\end{equation}
for any $Q^2$ and $Q_0^2$. Similar relation exists for the structure
function $g_2$ and $g_1+g_2$. \\

The sum rule (9) depends strongly on the small $x$ behavior of the
structure function $g_1$. The Regge theory predicts as 
$g_1\sim \beta x^{-\alpha_i(0)}$ with $\alpha_i(0) \leq 0$
where $i$ denotes $a_1$ and $f_1$ trajectory. In this case,
the sum rule is convergent except at $\alpha_i=0$.
The extrapolation of the DGLAP fit to the unmeasured small $x$
region have large ambiguity\cite{grsv}.
The double logarithmic $(log (1/x))^2$ resummation 
give more singular behavior than the Regge
theory\cite{bade}. The latter cases shows the sum rule (9) is divergent.
Though, whether the sum rule diverges or not can not be judged
rigorously by these discussions, in view of these situations,
it is important to discuss the regularization of the sum rule 
and give it a physical meaning even when the sum rule is divergent. 
Now,the formally divergent sum rule of the forward direction 
in the null-plane formalism was known to be regularized 
by the analytical continuation of the sum rule in
the non-forward direction\cite{dealwis,DP}. This method was further
developed to the current anticommutation relation on the
null-plane in Ref.\cite{kore84,kore93,kore98}. 
We consider the non-forward matrix element corresponding
to the reaction 'current($q_1$)+nucleon($p_1$) $\to$
current($q_2$)+nucleon($p_2$)', where we define
$K=(q_1+q_2)/2, P=(p_1+p_2)/2, \Delta = q_2 - q_1, \Delta^2=t,
q_1^2=q_2^2, p_1^2=p_2^2,$ and 
$S^{\mu}=\bar{(p_2s_2)}\gamma^{\mu}\gamma^5u(p_1s_1)/2$.
The explicit expression of the matrix element of the 
current anticommutation relation on the null-plane was given for the case 
$\mu =+,\nu =+$. The same reasoning can be done for
the case $\mu =+,\nu =i$. Since we need kinematics of the
spin-dependent part in doing this, we explain it.
The spin-dependent part for the conserved vector current has been known
to be expressed by the 13 structure functions.\cite{DP}
In case of the $q_1^2=q_2^2$ and $p_1^2=p_2^2$
in this paper, 5 structure functions becomes zero
under the time-reversal invariance. Among the remaining
8 structure functions, the tensor structure of 6 structure
functions are proportional to $\Delta$. Hence only 2
structure functions are left. We can take these two
structure functions as the ones which exactly become
$\tilde{G}_1$ and the $\tilde{G}_2$ in the forward
matrix element and separate
out the terms which remain in the forward limit.
Now, under this kinematical structure, 
since $P,K,\Delta$ are independent variables
we obtain the sum rules of the same forms as Eqs.(7)-(9)
in the non-forward case. Each quantity which appears
in the sum rules (7)-(9) is replaced by the quantity
in the non-forward one(see Eq.(2.11) and Eq.(4.1) in Ref.\cite{kore84}).
Then by assuming a moving pole or cut, we analytically
continue them to the forward direction. 
Since the sum rules take the same form as the forward ones,
we can effectively do this manipulation by using the
sum rules in the forward direction by introducing
the parameter which reflects the moving pole or cut.
In case of the sum rule (9), we rewrite it as
\begin{eqnarray}
\lefteqn{\int_0^1\frac{dx}{x}\{ g_1(x,Q^2) - f(x,Q^2)\} +
 \int_0^1\frac{dx}{x}f(x,Q^2)}&&\nonumber \\
&=& \int_0^1\frac{dx}{x}\{ g_1(x,Q_0^2) - f(x,Q_0^2)\} +
 \int_0^1\frac{dx}{x}f(x,Q_0^2),
\end{eqnarray}
where $f(x,Q^2)$ is the term which includes a possible
divergent piece in $g_1(x,Q^2)$. Let us consider the
simple pole case as $\alpha (0,\epsilon )=a - \epsilon$,
and set $f(x,Q^2) = \beta (Q^2)x^{-\alpha (0,\epsilon)} + f_1(x,Q^2)$.
The $\epsilon$ is a parameter which reflects the moving pole and
proportional to $t$ in the non-forward case.
The cases where more moving poles which give divergent behavior
in the forward exist can be done simply
by repeating the argument below with a minor trivial modification.
We first take $\epsilon > a$ and obtain
\begin{equation}
\int_0^1\frac{dx}{x}f(x,Q^2) = \frac{\beta (Q^2)}{\epsilon -a}
+\int_0^1\frac{dx}{x}f_1(x,Q^2),
\end{equation}
where the integral over $f_1$ on the right hand side of Eq.(11)
is finite in the limit $\epsilon \to a$ and $\epsilon \to 0$.
Then we take out the pole from both sides of Eq.(10) by obtaining
the condition $\beta (Q^2)=\beta (Q^2_0)$, and take the limit
$\epsilon \to 0$. Thus we obtain
\begin{eqnarray}
\lefteqn{\int_0^1\frac{dx}{x}\{ g_1(x,Q^2) - f(x,Q^2)\}}&&\nonumber \\
&=& \int_0^1\frac{dx}{x}\{ g_1(x,Q_0^2) - f(x,Q_0^2)\} +
 \int_0^1\frac{dx}{x}\{f(x,Q_0^2) -f(x,Q^2)\},
\end{eqnarray}
where we have replaced the integral over $f_1$ to $f$ in the
final result under the recognition that the coefficient of
a possible divergent piece in $g_1$ is $Q^2$ independent.
Practically, we do not care about this condition
since it is necessary only in the $x\to 0$ limit.
In this sense, as far as we can find a large
cancellation in the high energy region, Eq.(12)
can be considered to be valid.\\

Now, let us first consider the results of the sum rule (9) 
for the proton target when it is convergent.
In the sum rule, $g_1$ in general includes
the elastic contribution.  Since our concern here
lies in the behavior of $g_1$ in the low $Q^2$ region
and we take $Q_0^2=0$ on the right side of Eq.(9), 
we keep the Born contribution on both sides of Eq.(9)\cite{com}, and obtain
\begin{equation}
\int_0^1\frac{dx}{x}g_1^{p}(x,Q^2)= B(Q^2)
-\frac{1}{8\pi^2\alpha_{em}}\int_{\nu_0}^{\infty}d\nu\{
\sigma_{3/2}^{\gamma p} - \sigma_{1/2}^{\gamma p}\},
\end{equation}
with
\begin{equation}
B(Q^2)=\frac{1}{2}\{F_1^{p}(0)(F_1^{p}(0)+F_2^p(0)) -
F_1^{p}(Q^2)(F_1^{p}(Q^2)+F_2^p(Q^2)) \},
\end{equation}
where we use the relation at $Q^2=0$
\begin{equation}
G_1^p(\nu)=\frac{-1}{8\pi^2\alpha_{em}}\{\sigma_{3/2}^{\gamma p} - \sigma_{1/2}^{\gamma p}\}.
\end{equation}
Here the Dirac form factor takes the values $F_1^p(0)=1$ and
$F_2^p(0)=1.79$, and we give the nucleon mass dependence
explicitly.

Now $B(Q^2)$ is well known experimentally. We plot it in
Fig.1 by using the standard dipole fit\cite{Arn}
\begin{equation}
G_E^p=\frac{1}{(1+\frac{Q^2}{0.71})^2}, 
G_M^p=\mu_pG_E^p,
\end{equation}
where the anomalous magnetic moment $\mu_p=2.793$. 
The relation between the Dirac form
factor and the Sacks ones are $G_M^p=F_1^p+F_2^p$
and $\displaystyle{G_E^p=F_1^p - \frac{Q^2}{4m_N^2}F_2^p}$.\\
\begin{figure}
\includegraphics[width=100mm]{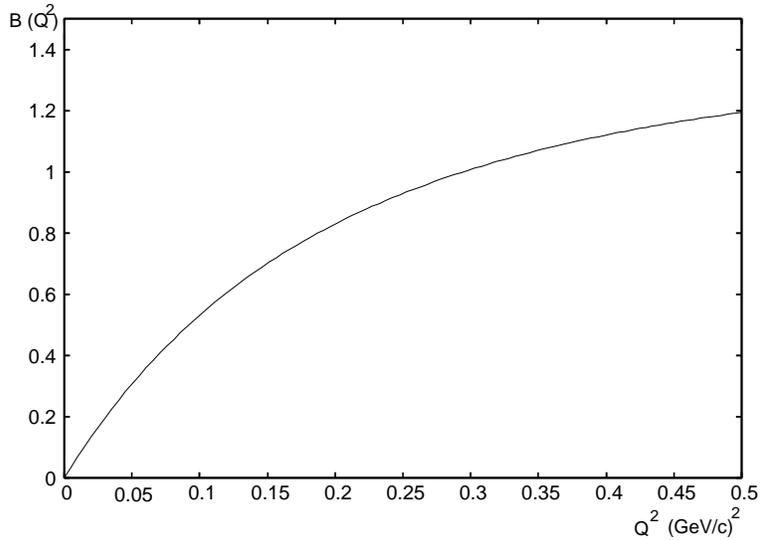}
\caption{The difference of the Born terms as given by the $B(Q^2)$}
\end{figure}
\mbox{}\\

Let us turn to the estimate of the integral of the cross
section of the photo-production and rewrite the sum rule (13)
by applying Eq.(12). Recently,the measurement of the
$\Delta \sigma =\sigma_{3/2}-\sigma_{1/2}$ was
reported\cite{ahrens,dutz}. According to these,
we can estimate the integral on the right side of
Eq.(13) up to $E\sim 2$ GeV directly with use of the experimental
value where $E$ is the energy of the photon in the laboratory frame. 
The contribution above this comes both from the resonances and 
the non-resonant terms. Though the contribution from the former is
small, the one from the latter is expected to be very large.
On the other hand, to estimate the left side of Eq.(13),
we need information of the $g_1$ in the very small $x$ region
which is also expected to give a large contribution.
In the small $Q^2$ region, if we take a sufficiently
large energy, high energy behavior of the total cross
section of the photo-production may coincide with
that of the $g_1$ with exactly the same proportional
constant as given in Eq.(15).
In fact, there is a phenomenological parameterization which has this
property\cite{Sim}. Then, the possible large
contributions from both sides of Eq.(13) may cancel out.
This is a situation where the regularized sum rule (12)
can be used. 
Thus, by setting $\nu =m_NE$ in the laboratory frame,
for arbitrary $Q^2$, we equate $f(x,Q^2)$ in the sum rule (12) 
as $g_1(x,Q^2)$ below $x=x_c$ and 0 above it where  $x_c=Q^2/2\nu_c^Q$,$\nu_c=m_NE_c$, and
$\nu_c^Q=\nu_c+Q^2/2$ with  $E_c=2$ GeV. Then we divide the
integral from $0$ to $x_c(Q^2)$ and $x_c(Q^2)$ to $1$ for the $g_1(x,Q^2)$
and $f(x,Q^2)$ and from $0$ to $x_c(Q_0^2)$ and $x_c(Q_0^2)$ to $1$
for the $g_1(x,Q_0^2)$ and $f(x,Q_0^2)$. Using the fact that the
$g_1(x,Q^2)=f(x,Q^2)$ below $x=x_c(Q^2)$ and $g_1(x,Q_0^2)=f(x,Q_0^2)$ 
below $x=x_c(Q_0^2)$, we can rewrite the sum rule (12).
Then, by taking the $Q_0^2=0$ and using the relation
(15), the sum rule(12) where the Born term is separated 
out is given as
\begin{equation}
\int_{x_c}^1\frac{dx}{x}g_1^{p}(x,Q^2)= B(Q^2)
-\frac{1}{8\pi^2\alpha_{em}}\int_{\nu_0}^{\nu_c}d\nu\{
\sigma_{3/2}^{\gamma p} - \sigma_{1/2}^{\gamma p}\}+K(E_c,Q^2),
\end{equation}
where
\begin{equation}
K(E_c,Q^2)=\frac{1}{8\pi^2\alpha_{em}}\int_{\nu_c}^{\infty}d\nu\{
\sigma_{1/2}^{\gamma p} - \sigma_{3/2}^{\gamma p}\}
-\int_{\nu_c^Q}^{\infty}\frac{d\nu}{\nu}g_1^{p}(x,Q^2).
\end{equation}
Here the integral over $\nu$ in Eq.(18) should be
understood to be done after we subtract the high 
energy behavior of both the photoproduction and the $g_1^p(x,Q^2)$.
The sum rule (17) is the regularized version of the sum rule (13),
where the high energy contribution is subtracted out.
We take the $g_1^{p}$ in the $K(E_c,Q^2)$ as
the non-resonant contribution $g_1^{non-res.}$ in Ref.\cite{Sim}.
We neglect the resonant contribution above $E_c$, since
inclusion of these contribution does not affect the following
discussions. We further approximate the $g_1^{non-res.}$ as 
$g_1^{non-res.}=g^{\Delta\sigma}$ where
$g^{\Delta\sigma}$ is the contribution arising from the
transverse asymmetry $A_1(x,Q^2)$ and also defined in Ref.\cite{Sim}.
The approximation here is equivalent to neglect $g_2$
in the transverse asymmetry, and its effect is negligible
in the evaluation of the $K$ above $E_c=2$GeV. Further, for $Q^2 < 0.1$(GeV/c)$^2$,
we cut the integral in $K$ at $E=100$GeV since the integrand can almost
be regarded as zero in this energy region. Under these approximations, 
for example,we obtain $K(2,0.05)\sim -0.014$ and $K(2,0.1)\sim -0.027$.\cite{para} 
Thus there is a large cancellation here.
On the other hand, using the experimental data given in Ref.\cite{dutz}, we obtain
\begin{equation}
\frac{m_N}{8\pi^2\alpha_{em}}\int_{E_0}^{2}dE\{
\sigma_{3/2}^{\gamma p} - \sigma_{1/2}^{\gamma p}\}
\sim 0.45.
\end{equation}
By calculating $K(2,Q^2)$ for each $Q^2$,  we plot
the left-hand side of Eq.(17) as a function of $Q^2$ in Fig.2.
The dotted curve (a) is the one where the contribution from
$K(2,Q^2)$ is neglected and the curve (b) is the one
where the contribution from $K(2,Q^2)$ is included. From it
we find that the integral
\begin{equation} 
\int_{x_c}^1\frac{dx}{x}g_1^{p}(x,Q^2)
\end{equation}
become zero in the region near $Q^2\sim 0.09$(GeV/c), where $x_c=0.024$.
\begin{figure}
\includegraphics[width=100mm]{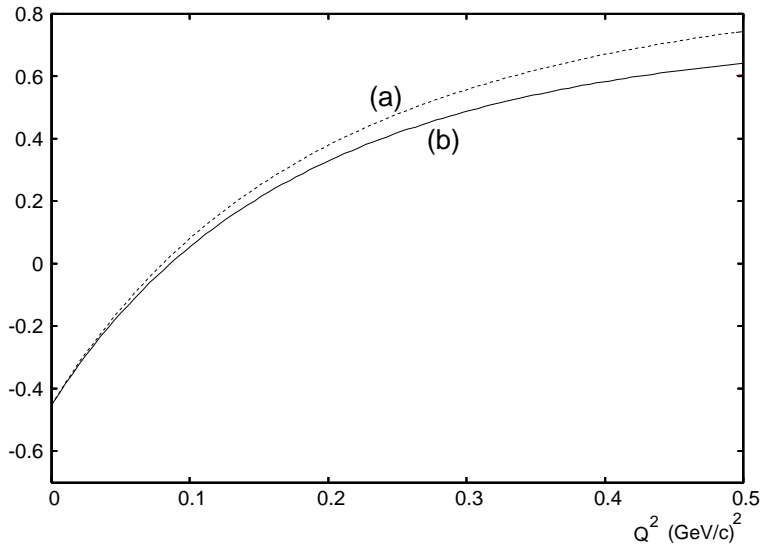}
\caption{The estimate of the left-hand side of Eq.(17).
The dotted curve (a) is the one given by the
Born contribution $B(Q^2)$ and Eq.(19) by neglecting $K(2,Q^2)$.
The curve (b) is the one including the contribution from $K(2,Q^2)$. }
\end{figure}
Now we can use the parameters in Ref.\cite{Sim} not only to show
the smallness of the $K(E_c,Q^2)$ at small $Q^2$ but also
to check the right hand side of Eq.(19). Since the value is slightly
different from 0.45 as given in Eq.(19), we find that
the zero of Eq.(20) occurs at $Q^2\sim 0.16$(GeV/c)$^2$ with $x_c=0.042$.
This zero point is a little bit larger than $Q^2\sim 0.09$(GeV/c)$^2$.
However, in this model, we see that this same rapid change of the
inelastic contribution of the $g_1^p$ gives the sign change of the
generalized Drell-Hern-Gerasimov sum rule.\\

In summary, we have shown that the appropriately defined moment at
$n=0$ of the polarized structure function $g_1^p$ defined at the left
hand side of Eq.(17) becomes zero at small $Q^2$ near
$Q^2\sim0.09$(GeV/c)$^2$, and that the sign change occurs at this point.
Since the Ellis-Jaffe sum rule corresponds to the moment at $n=1$
which is more sensitive to the low energy behavior than the present
one at $n=0$, the negative resonance contribution is
enhanced in it. Therefore the fact that the moment at $n=0$ change sign
shows that the sign of the Drell-Hern-Gerasimov sum rule
is opposite to that of the Ellis-Jaffe sum rule. The origin
of the sign change is the rapid change of the Born term at low $Q^2$
which is compensated by the inelastic contribution. Thus the fact that
the rapid change of the Born term is below $Q^2<0.5$ (GeV/c)$^2$
explains why the sign change occurs at very small $Q^2$.
The compensation is the reflection of the $Q^2$ independence
of the moment at $n=0$ as given by the sum rule (9), from which
the sum rule (17) is derived. Phenomenological importance of the sum
rule (17) lies in the fact that we can investigate the $Q^2$ dependence of the 
resonance structure in the low and the intermediate energy 
region at low $Q^2$ without worrying about the correction
from the high energy behavior. 
Now, the sum rule (17) can be used for any $Q^2$. For example,
it can be used for the $Q^2$ in the deep
inelastic region. In this case, the $Q^2$ dependent piece in the Born
terms rapidly become 0 and hence it can be neglected.
On the other hand, we get a large contribution from $K(Q^2,E_c)$,
if we take $E_c=2$ GeV. 
We need more data in the small $x$ region
together with information of the photo-production to see
how far the high energy behavior is canceled. If we can
find a large cancellation, we
take a large $E_c$ such that a contribution from
$K(Q^2,E_c)$ becomes small. In this way, we can extend the
analysis of the sum rule (17) to the larger $Q^2$ region
where the resonance contribution turns to the continuum
contribution and study their relation.


\begin{references}
\bibitem{com}In the previous paper\cite{kore}, we take $Q^2$ in the
deep inelastic region and $Q_0^2$ as 0. Since the Born term
becomes negligible in the deep inelastic region, it
was kept only on the right side of Eq.(9). Further $g_I$ in the
paper\cite{kore} was two times larger than the usually defined ones
given in this paper.
\end{references}
\end{document}